\documentclass[epj]{webofc}
\usepackage[varg]{txfonts}   
\usepackage{graphicx,color} 
\usepackage{bm}       
\usepackage{amsmath}  
\usepackage{amssymb}
\woctitle{21st International Conference on Few-Body Problems in Physics}
\begin{document}
\title{Quark model with chiral-symmetry breaking\\ and confinement in the Covariant Spectator Theory}
\author{Elmar P. Biernat\inst{1}\fnsep\thanks{\email{elmar.biernat@tecnico.ulisboa.pt}} \and        
        M. T. Pe\~{n}a\inst{1,2}
        \and 
        J. E. Ribeiro\inst{3} \and
        A. Stadler\inst{4,1}\and 
        F. Gross\inst{5}      
}
\institute{CFTP, Instituto Superior T\'ecnico, Universidade de Lisboa, 1049-001 Lisboa, Portugal 
\and Departamento de F\'isica, Instituto Superior T\'ecnico, Universidade de Lisboa, 1049-001 Lisboa, Portugal 
\and CeFEMA, Instituto Superior T\'ecnico, Universidade de Lisboa, 1049-001 Lisboa, Portugal
\and Departamento de F\'isica, Universidade de \'Evora, 7000-671 \'Evora, Portugal
\and Thomas Jefferson National Accelerator Facility (JLab), Newport News, Virginia 23606, USA          
          }

\abstract{
  We propose a model for the quark-antiquark interaction in Minkowski space using the Covariant Spectator
Theory. We show that with an equal-weighted scalar-pseudoscalar structure for the confining part of our interaction kernel the axial-vector Ward-Takahashi identity is preserved and our model complies with the Adler-zero constraint for $\pi$-$\pi$-scattering imposed by chiral symmetry. 
}
\maketitle
\section{Introduction}
\label{intro}

As the lightest quark-antiquark bound state the pion is of particular importance for our understanding of confinement and spontaneous chiral-symmetry breaking (S$\chi$SB). It emerges non-perturbatively from the strong interaction and it is at the same time identified with the Goldstone boson associated with S$\chi$SB. Various modern approaches have addressed the non-perturbative dynamics underlying such hadronic systems. For instance, lattice-QCD simulations~\cite{Edwards,Guo}, light-front quantum field theory~\cite{Brodsky:1997de,Carbonell:1998rj}, as well as models based on the Dyson-Schwinger--Bethe-Salpeter (DSBS) approach and the mass gap equation~\cite{Bicudo:1989sj,Nefediev:2004by,Alkofer:2000wg,Maris:2003vk,Fischer:2006ub,Rojas:2013tza} have significantly contributed to an understanding of a wide range of hadronic phenomena. The framework we use is the Covariant Spectator Theory (CST)~\cite{Gro69,Gross:1991te,Gross:1991pk,Gross:1994he,Savkli:1999me,PhysRevD.89.016005} --- another modern field-theoretic approach that implements S$\chi$SB through the famous Nambu--Jona-Lasinio mechanism, similarly to DSBS. Whereas the latter is usually treated in a Euclidean formulation, CST is established in Minkowski space, an advantage for instance when computing form factors in the timelike region. Another distinct feature of CST is its capability of accommodating a Lorentz-scalar confining interaction kernel without destroying chiral symmetry, which is of particular importance in view of the approaches~\cite{Michael:1985rh,Allen:2000sd,Bali:1997am} suggesting the existence of a scalar component for the quark-antiquark interaction. In the present work we study to what extent such confining forces can be made consistent with S$\chi$SB. Our strategy is to start from the most general Lorentz structure for the interaction kernel and then determine the constraints imposed by chiral symmetry, similarly to what has been done in Ref.~\cite{PhysRevD.47.1145} for a different formalism. It turns out that a CST model with scalar confinement, together with an equal-weighted pseudoscalar counterpart satisfies the S$\chi$SB condition of the Adler consistency zero~\cite{Adler_PhysRev.137.B1022} in $\pi$-$\pi$ scattering in the chiral limit.
\section{Axial-vector Ward-Takahashi identity}
\label{sec-1}
Explicit and spontaneous chiral-symmetry breaking is expressed in quantum field theory through the axial-vector Ward-Takahashi identity (AV-WTI) involving the dressed quark propagators and the dressed axial-vector and pseudoscalar vertices. When dealing with strong quark form factors according to Gross and Riska~\cite{Gro87,Gro93,Gro96} the AV-WTI reads
\begin{eqnarray}
P_\mu \Gamma^{5\mu}_{R}(p',p)+2m_0 \Gamma^5_R(p',p)=\tilde S^{-1} (p')\gamma^5+\gamma^5\tilde S^{-1} (p)\,,\label{eq:AVWTI}
\end{eqnarray}
 where $\Gamma^{5\mu}_R(p',p)$ and  $\Gamma^5_R(p',p)$ are the dressed axial-vector and pseudoscalar vertices, respectively, $m_0$ is the bare quark mass, $p$ and $p'$ are the incoming and outgoing quark momenta, respectively, $P=p'-p$ is the momentum flowing into the vertex, and $\tilde S (p)$ is the dressed quark propagator as introduced in Ref.~\cite{PhysRevD.90.096008}. The combination on the LHS of Eq.~(\ref{eq:AVWTI}) is sometimes called the dressed axial vertex $\Gamma^{A}_R (p^\prime,p)$, which is the solution of an inhomogeneous CST Bethe-Salpeter equation (CST-BSE),
 \begin{eqnarray} 
\Gamma^{A}_R (p^\prime,p)=\gamma^{A}_R (p^\prime,p)+\mathrm i
\int_{k0} \mathcal V_R(p-k) 
\tilde S(k') \Gamma^{A}_R (k',k) \tilde S(k) \,, \qquad  \label{eq:CSTBSEGammaA}
\end{eqnarray}
 where $\gamma^{A}_R (p^\prime,p)$ is the bare axial vertex, $\mathcal V_R(p-k)$ is the covariant $q\bar q$ interaction kernel depending only on the four-momentum transfer $p-k=p'-k'$, and \lq\lq $k0$'' indicates the charge-conjugation invariant CST prescription for performing the $k_0$ contour integration~\cite{Savkli:1999me}. The most general structure of the  CST generalization of the linear-confining potential, together with a vector--axial-vector remainder, is given by
 \begin{eqnarray}
 \mathcal V_R(p-k)&=& V_{LR}(p-k)\Big[\lambda_S ({\bf 1}\otimes {\bf 1})+\lambda_P (\gamma^5\otimes\gamma^5) +\lambda_V
( \gamma^\mu 
\otimes \gamma_{\mu})+\lambda_A (\gamma^5\gamma^{\mu} \otimes \gamma^5\gamma_{\mu}) \nonumber\\&&+
\frac{\lambda_T}{2}(\sigma^{\mu\nu}\otimes\sigma_{\mu\nu})\Big] +V_{CR}(p-k)
\Big[\kappa_V(\gamma^\mu\otimes\gamma_\mu)+\kappa_A (\gamma^5\gamma^{\mu} \otimes \gamma^5\gamma_{\mu})\Big]
\,,\label{eq:kernel} 
\end{eqnarray}
where $V_{LR}$ and $V_{CR}$ are the Lorentz-invariant momentum-dependent parts of the linear-confining and remaining kernels, respectively. The corresponding weight parameters $\lambda_i$ and $\kappa_i$  [with $i=S$ (scalar), $P$ (pseudoscalar), $V$ (vector), $A$ (axial-vector), and $T$ (tensor)] are arbitrary constants. Further, $V_{LR}$ satisfies the CST generalization of the non-relativistic condition $V_L(r=0)=0$, given by
\begin{eqnarray}
 \int \frac{\mathrm d^3 k}{E_k} V_{LR} (p\pm\hat k)=0\,,
 \label{eq:VLzero}
\end{eqnarray}
 where $E_k=\sqrt{m^2+\vec k^2}$, $\hat k=(E_k,\vec k)$, and $m$ is the dressed quark mass. For this kernel with $\lambda_S=\lambda_P$ it has been shown~\cite{PhysRevD.90.096008} that the AV-WTI (\ref{eq:AVWTI}) together with the CST-BSE (\ref{eq:CSTBSEGammaA}) implies that $\tilde S (p)$ is the solution of the CST-Dyson Equation (CST-DE),
\begin{eqnarray}
 \tilde S^{-1} (p)=\tilde S_0^{-1} (p)-\mathrm i\int_{k0} 
{\cal V}_R(p-k) \tilde S(k)\, ,
\label{eq:CST-DE2}
\end{eqnarray}
where $\tilde S_0$ is the bare quark propagator, which obeys an AV-WTI involving $\gamma^{A}_R (p^\prime,p)$. It turns out that $\gamma^{A}_R (p^\prime,p)$ vanishes in the chiral limit of vanishing bare quark mass, $m_0\rightarrow 0$, and vanishing vertex momentum, $P\rightarrow 0$.  In this limit, the CST-BSE~(\ref{eq:CSTBSEGammaA}) becomes homogeneous and identical to the zero-mass pion CST equation for the pion vertex function in the chiral limit, $\Gamma^\pi_{R\chi}$, which implies the relation
\begin{eqnarray}
\Gamma^{A}_{R\chi}(p,p) \propto\Gamma^\pi_{R\chi}(p,p)\,.
\label{eq:Gammachi}
\end{eqnarray}
Because of the condition~(\ref{eq:VLzero}), and imposing equal weights to the scalar and pseudoscalar terms in the interaction kernel only $\mathcal V_{CR}$ contributes to the chiral-limit pion equation and to the scalar part of the CST-DE~\eqref{eq:CST-DE2}. This corresponds to dynamical quark mass generation. Therefore, the linear-confinement part $\mathcal V_{LR}$ that includes the scalar, pseudoscalar and tensor structures in our model, entirely decouples from these equations~\cite{Gross:1991pk}. 
\section{$\pi$-$\pi$ scattering and Adler zero}
\begin{figure}
\centering
\includegraphics[width=13cm,clip]{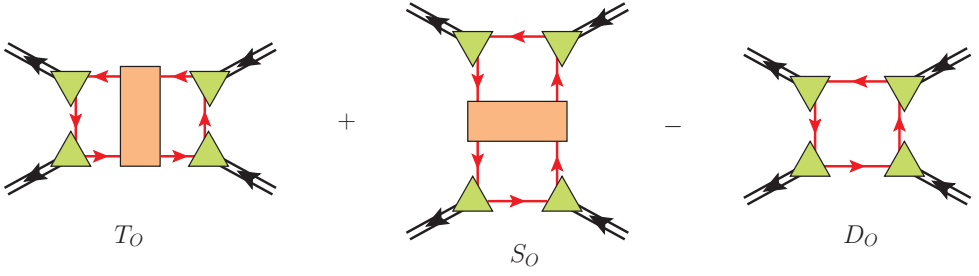}
\caption{The $O$ contributions to $\pi$-$\pi$~scattering. The orange boxes denote the unamputated quark-quark scattering amplitudes.}
\label{fig:pipiscattering_O}       
\end{figure}

\begin{figure}
\centering
\includegraphics[width=13cm,clip]{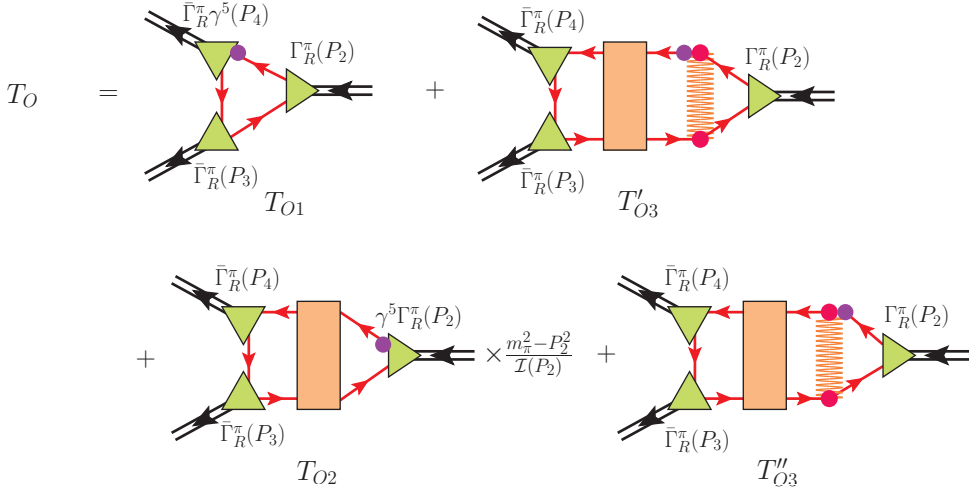}
\caption{The $T_O$ term. A purple blob denotes a $\gamma^5$ and $\mathcal I(P)$ is a non-vanishing normalization integral. }
\label{fig:term1_2}       
\end{figure}

Consistency with chiral symmetry implies the vanishing of the $\pi$-$\pi$~scattering amplitude in the chiral limit~\cite{Adler_PhysRev.137.B1022}. This property, known as the Adler zero, is due to remarkable cancellations, which occur between different scattering diagrams that go beyond the (lowest-order) impulse approximation. In these diagrams intermediate-state interactions to all orders are included through the complete quark-quark ladder sum~\cite{PhysRevD.65.076008,PhysRevC.67.035201}, which involves nine different types of contributions. The proof of the cancellations between these contributions is rather lengthy. To illustrate the technical procedure we discuss here only the diagrams shown in Fig.~\ref{fig:pipiscattering_O}. The same treatment applies to the other diagrams that have to be taken into account. The full proof is given in detail in Ref.~\cite{PhysRevD.90.096008}. 
In order to show that the sum of these diagrams vanishes in the chiral limit, one inserts an additional ladder sum at one pion vertex by using the spectral decomposition of the ladder sum. Then, by Eq.~(\ref{eq:Gammachi}), $\Gamma^\pi_{R\chi}$ is replaced by $\Gamma^{A}_{R\chi}$, allowing the application of the AV-WTI~\eqref{eq:AVWTI} between two ladder sums. For $T_O$ this results in 4 terms as depicted in Fig.~\ref{fig:term1_2}.
Because the sum $T_{O3}'+T_{O3}''$ is proportional to the anticommutator $\lbrace\mathcal V_R,\gamma^5\rbrace$, all vector and axial-vector contributions cancel. The scalar, pseudoscalar, and tensor structures from the confining potential integrate to zero in the chiral limit because of the decoupling property discussed in the previous section. The $T_{O2}$ term vanishes as consequence of the fact that the pion does not couple to the scalar channel. Finally, the $T_{O1}$ term, together with $S_{O}$ cancel exactly the $D_O$ term by applying again the AV-WTI.

\begin{acknowledgement}
This work received financial support from Funda\c c\~ao para a Ci\^encia e a 
Tecnologia (FCT) under Grants No.~PTDC/FIS/113940/2009 and No. CFTP-FCT (PEst-OE/FIS/U/0777/2013). The research leading to these results has received funding from the European Community's Seventh Framework Programme FP7/2007-2013 under Grant Agreement No.\ 283286. This work was also partially supported by Jefferson Science Associates, LLC, under U.S. DOE Contract No. DE-AC05-06OR23177. 
\end{acknowledgement}

%

\begin{thebibliography}{27}

\bibitem{Edwards}
R.G. Edwards, N.~Mathur, D.G. Richards, S.J. Wallace, Phys. Rev. D \textbf{87},
  054506 (2013)

\bibitem{Guo}
P.~Guo, J.J. Dudek, R.G. Edwards, A.P. Szczepaniak, Phys. Rev. D \textbf{88},
  014501 (2013)

\bibitem{Brodsky:1997de}
S.J. Brodsky, H.C. Pauli, S.S. Pinsky, Phys. Rept. \textbf{301}, 299 (1998)

\bibitem{Carbonell:1998rj}
J.~Carbonell, B.~Desplanques, V.~Karmanov, J.~Mathiot, Phys. Rept.
  \textbf{300}, 215 (1998)

\bibitem{Bicudo:1989sj}
P.J. de~A.~Bicudo, J.E.F.T. Ribeiro, Phys. Rev. D \textbf{42}, 1635 (1990)

\bibitem{Nefediev:2004by}
A.V. Nefediev, J.E.F.T. Ribeiro, Phys. Rev. D \textbf{70}, 094020 (2004)

\bibitem{Alkofer:2000wg}
R.~Alkofer, L.~von Smekal, Phys. Rept. \textbf{353}, 281 (2001)

\bibitem{Maris:2003vk}
P.~Maris, C.D. Roberts, Int. J. Mod. Phys. E \textbf{12}, 297 (2003)

\bibitem{Fischer:2006ub}
C.S. Fischer, J. Phys. G \textbf{32}, R253 (2006)

\bibitem{Rojas:2013tza}
E.~Rojas, J.~de~Melo, B.~El-Bennich, O.~Oliveira, T.~Frederico, J. High Energy
  Phys. \textbf{1310}, 193 (2013)

\bibitem{Gro69}
F.~Gross, Phys. Rev. \textbf{186}, 1448 (1969)

\bibitem{Gross:1991te}
F.~Gross, J.~Milana, Phys. Rev. D \textbf{43}, 2401 (1991)

\bibitem{Gross:1991pk}
F.~Gross, J.~Milana, Phys. Rev. D \textbf{45}, 969 (1992)

\bibitem{Gross:1994he}
F.~Gross, J.~Milana, Phys. Rev. D \textbf{50}, 3332 (1994)

\bibitem{Savkli:1999me}
C.~Savkli, F.~Gross, Phys. Rev. C \textbf{63}, 035208 (2001)

\bibitem{PhysRevD.89.016005}
E.P. Biernat, F.~Gross, M.T. Pe\~na, A.~Stadler, Phys. Rev. D \textbf{89},
  016005 (2014)

\bibitem{Michael:1985rh}
C.~Michael, Phys. Rev. Lett. \textbf{56}, 1219 (1986)

\bibitem{Allen:2000sd}
T.J. Allen, M.G. Olsson, S.~Veseli, Phys. Rev. D \textbf{62}, 094021 (2000)

\bibitem{Bali:1997am}
G.S. Bali, K.~Schilling, A.~Wachter, Phys. Rev. D \textbf{56}, 2566 (1997)

\bibitem{PhysRevD.47.1145}
J.E. Villate, D.S. Liu, J.E. Ribeiro, P.J. de~A.~Bicudo, Phys. Rev. D
  \textbf{47}, 1145 (1993)

\bibitem{Adler_PhysRev.137.B1022}
S.L. Adler, Phys. Rev. \textbf{137}, B1022 (1965)

\bibitem{Gro87}
F.~Gross, D.O. Riska, Phys. Rev. C \textbf{36}, 1928 (1987)

\bibitem{Gro93}
F.~Gross, Y.~Surya, Phys. Rev. C \textbf{47}, 703 (1993)

\bibitem{Gro96}
Y.~Surya, F.~Gross, Phys. Rev. C \textbf{53}, 2422 (1996)

\bibitem{PhysRevD.90.096008}
E.P. Biernat, M.T. Pe\~na, J.E. Ribeiro, A.~Stadler, F.~Gross, Phys. Rev. D
  \textbf{90}, 096008 (2014)

\bibitem{PhysRevD.65.076008}
P.~Bicudo, S.~Cotanch, F.~Llanes-Estrada, P.~Maris, E.~Ribeiro, A.~Szczepaniak,
  Phys. Rev. D \textbf{65}, 076008 (2002)

\bibitem{PhysRevC.67.035201}
P.~Bicudo, Phys. Rev. C \textbf{67}, 035201 (2003)

\end{thebibliography}
%
%

%

\end{document}